\documentclass[english]{sbrt}
\usepackage[english]{babel}
\usepackage[utf8]{inputenc}

\newcommand{\jump}[1]{{\vspace{20pt}}}

\usepackage{graphicx} 
\usepackage{float}

\usepackage[nolist]{acronym}

\usepackage{xcolor}
\usepackage[normalem]{ulem}

\usepackage[caption=false,font=footnotesize]{subfig}

\usepackage{url}

\usepackage{makecell}

\newcommand{\SecRef}[2][]{Section#1~\ref{#2}}
\newcommand{\FigRef}[2][]{Fig.#1~\ref{#2}}

\newcommand{\TabRef}[2][]{Table#1~\ref{#2}}

\begin{document}

\title{Experimental comparison of 5G SDR platforms: \\srsRAN x OpenAirInterface}

	\author{Ruan P. Alves, João Guilherme A. da S. Alves, Mikael R. Camelo, \\ Wilker O. de Feitosa, Victor F. Monteiro, Fco. Rodrigo P. Cavalcanti
		\thanks{The authors are with the Wireless Telecommunications Research Group (GTEL), Federal University of Cear\'{a} (UFC), Fortaleza, Cear\'{a}, Brazil. The work of Victor F. Monteiro was supported by CNPq under Grant 308267/2022-2.}%
	}

	\maketitle

	\markboth{XLI BRAZILIAN SYMPOSIUM ON TELECOMMUNICATIONS AND SIGNAL PROCESSING - SBrT 2023, OCTOBER 08--11, 2023, S\~{A}O JOSÉ DOS CAMPOS, SP}{}

\begin{abstract}

A \ac{sdr} platform is a communication system that implements as software functions that are typically implemented in dedicated hardware. %
One of its main advantages is the flexibility to test and deploy radio communication networks in a fast and cheap way. %
In the context of the \ac{5G} of wireless cellular networks, there are open source \ac{sdr} platforms available online. %
Two of the most popular \ac{sdr} platforms are srsRAN and OpenAirInterface. %
This paper presents these two platforms, the characteristics of the networks created by them, the possibilities of changes in their interfaces and configurations, and also their limits. %
Moreover, in this paper, we also evaluate and compare both platforms in an experimental setup deployed in a laboratory. 

\end{abstract}

\begin{keywords}
Software-defined radio, srsRAN, OpenAirInterface, 5G.
\end{keywords}

\acresetall

%%%%%%%%%%%%%%%%%%%%%%%%%%%%%%%%%%%%%%%%%%%%%%%%%%%%%%%%%
\section{Introduction}
%%%%%%%%%%%%%%%%%%%%%%%%%%%%%%%%%%%%%%%%%%%%%%%%%%%%%%%%%
%Correções consideradas: 
% Acrônimos não definidos no início do texto ------- Feito.
% Mudar 100MHz para 100~MHz ------- Feito.
% Licenças de código aberto da SRSran e da OAI.  --------- On hold.
% Explicar porque OAI não é open-source (?) ----------- On hold.
% Tabela para as características principais  das plataformas (SEÇÃO II) ------- On hold.
% Explicar a parte de same reason (A razão são as adaptações para o sistema operacional.) --- Feito.
%  Explicar , na fig.4, porque a OAI possui quedas de throughtput (a rede possui uma limitação.)   ------ Feito.
% Explicar os resets da OAI na fig. 5 ----- Feito.
% to assembly and understand => to assemble and understand ------ FEITO.
% configure the COTS UE was easier since we only => configuring the COTS UE was easier since we only ---- FEITO.

Aiming at reducing costs and the time required for tests, deployment and updates, telecommunication operators and equipment vendors have been considering the replacement of network components typically implemented in dedicated hardware by software running in programmable computers~\cite{ref0}. %%are trying to replace parts of the network typically implemented in hardware by softwares. %
%One of the main enablers for this is the use of \ac{sdr}. %
One of the main enablers of this ``virtualization approach'' is the use of the concept of  \ac{sdr}~\cite{refb}. %
The ultimate goal of such approach is to decouple hardware and software which is expected to bring reduced costs and increased flexibility in network operation. %

A \ac{sdr} is a radio communication system where traditional hardware components, such as mixers, filters, amplifiers, and modulators/demodulators, are replaced or augmented by software processing. %
In a \ac{sdr}, the majority of the radio's functionality is implemented using software running on a general-purpose computer or embedded system. %

In simpler terms a \ac{sdr}-based network typically consists of: telecommunication radios, a software platform and computers. %
Regarding the telecommunication radios, they are responsible for transmission and reception of signals. %
Concerning the software platform, it implements as software the functions previously implemented as hardware, e.g., signal processing. %
Finally, the computers are connected to the telecommunication radios and are responsible for executing the software platform. %

Regarding the telecommunication radios, they can be found in different configurations and price ranges. %
They usually differ according to their capabilities, e.g., supported bandwidth, supported central frequency for transmission and reception, etc. %
Some popular devices used for this purpose are bladeRF, LimeSDR and \ac{usrp}. %

Concerning \ac{sdr} platforms that implement \ac{5G} networks, two prominent examples are \ac{srs} and \ac{oai}. % are two examples that can be cited. %
The \ac{srs} is an open source initiative, while the \ac{oai} is more closed. %
The \ac{oai} is older than the \ac{srs} and has a bigger and centralized team. %

%In this context, the present paper presents the results related to the implementation of two distinct \ac{5G} networks, each one using a different \ac{sdr} platform, i.e., \ac{srs} and \ac{oai}. %
In this context, the present paper extends our previous work using \acp{sdr}. %
On the one hand, in~\cite{refc}, we have deployed a \ac{4G} \ac{lte} network testbed using \acp{sdr} to evaluate an algorithm that predicts the signal quality of a link between an \ac{ue} and its serving \ac{enb}, i.e., a \ac{4G} base station. % 
On the other hand, the present paper presents the results related to the implementation of two distinct \ac{5G} networks, each one using a different \ac{sdr} platform, i.e., \ac{srs} and \ac{oai}. %

This paper is organized as follows. %
\SecRef{SEC:tech_info} presents a technical background related to both \acp{sdr} platform. %
The technical details related to our implementations are presented in \SecRef{SEC:methodology}. %
The obtained results and related discussions are presented in \SecRef{SEC:Results}. %
Finally, \SecRef{SEC:Conclusion} presents the conclusions of the present study and its future perspectives. %

%%%%%%%%%%%%%%%%%%%%%%%%%%%%%%%%%%%%%%%%%%%%%%%%%%%%%%%%%
\section{Background on 5G SDR-based Networks} \label{SEC:tech_info} 
%%%%%%%%%%%%%%%%%%%%%%%%%%%%%%%%%%%%%%%%%%%%%%%%%%%%%%%%%

\begin{figure}[t]
	\centering
	\includegraphics[width=\linewidth]{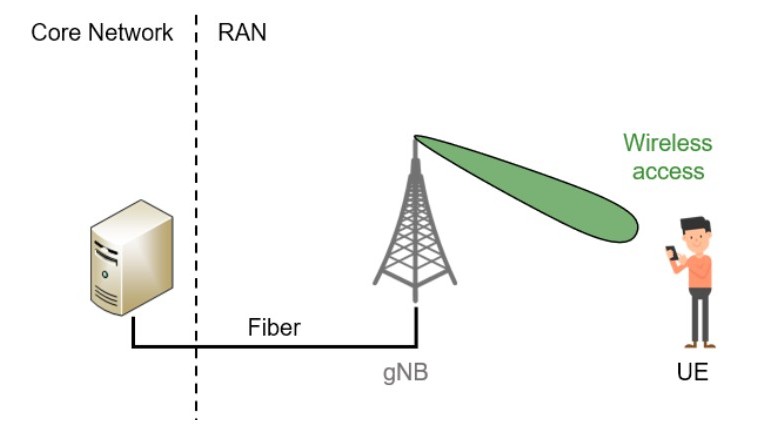}
	\caption{Simplified architecture of a 5G/NR network.}
	\label{Fig:netw-arch}
\end{figure}

The \ac{srs} is an end-to-end solution for \ac{lte} and \ac{nr} networks. %
Its \ac{nr} implementation has three main components: \textbf{\ac{srsue}}, \textbf{\ac{srsgnb}} and \textbf{\ac{srsepc}}, which implement, respectively, the three main elements of a \ac{nr} network (\FigRef{Fig:netw-arch}): %the \ac{ue}, the \ac{gnb} and the \ac{cn}, respectively: 
	
\begin{itemize}
	\item \textbf{\acl{ue}:} The \ac{ue} is any device used directly by an end-user to communicate. For example, a notebook and a smartphone.
	
	\item \textbf{\acl{gnb}:} it is the \ac{5G} \ac{bs}. It provides the connectivity between \ac{ue} and the \ac{cn}.	
	
	\item \textbf{\acl{cn}:} it is the heart of a \ac{5G}-\ac{nr} network, and ensures the efficient and reliable delivery of traffic between network nodes.
\end{itemize}

More specifically, the main characteristics of each \ac{srs} component are:

	\begin{itemize}

		\item \textbf{\ac{srsue}:} Support to: \ac{lte} and \ac{nr} (\ac{nsa}/\ac{sa}); \ac{fdd} and \ac{tdd} schemes; bandwidths: 1.4, 3, 5, 10, 15 and 20~MHz; up to 2x2 \ac{mimo}; up to 256 \ac{qam} modulation (only downlink); and physical \ac{sim} cards using a PC/SC reader.

		\item \textbf{srsgnb:} Support to: \ac{tdd} and \ac{fdd} schemes; subcarrier spacings: 15 and 30~kHz (FR1); bandwidths: 5, 10, 15, 20, 25, 30, 40, 50, 60, 70, 80, 90, 100~MHz (273 Physical Resource Blocks); \ac{mimo} up to 2x2; 64 \ac{qam} modulation in downlink.

		\item \textbf{\ac{srsepc}:} It is compatible only with \ac{lte}, and, moreover, supports only a small set of \ac{lte} features. For example, it does not support some features such as Inter-eNB (S1) roaming and VoLTE. To run these applications, it is necessary to use another \ac{cn} software.

	\end{itemize}

We highlight that with these components, \ac{srs} is able to run an end-to-end \ac{lte} and \ac{5G} \ac{nsa} network. %
For \ac{5G} \ac{sa}, it is necessary a separate \ac{cn}, since \textbf{\ac{srsepc}} only acts as an \ac{epc}, i.e., the \ac{4G} \ac{cn}. %
The  \ac{srs} Project documentation~\cite{ref5} suggests that if someone wants to deploy a complete \ac{5G} \ac{sa} network, he/she can replace its \ac{cn} by the \ac{cn} provided by the Open5GS, where the Open5GS is a project that provides only the \ac{cn}. % 
%Therefore,  as suggested by the \ac{srs} Project documentation \cite{ref5}, Open5GS was used as \ac{cn},
	
The second platform, \ac{oai}, is a mobile network solution, which, as the \ac{srs}, is designed to build end-to-end \ac{lte} and \ac{nr} networks. %
Despite the similarities, there are differences between these two platforms. %
For example, the nomenclature of the components. %
In \ac{oai} we have the \ac{ue} split into \textbf{nr-uesoftmodem} for \ac{nr} and \textbf{lte-uesoftmodem} for \ac{lte}; moreover we have \textbf{nr-softmodem} as \ac{gnb} for \ac{nr} and \textbf{lte-softmodem} as \ac{enb} for \ac{lte}; concerning the \ac{cn}, it has \textbf{\ac{oai5cn}} and \textbf{\ac{oaicn}}, which are the \ac{5G} and \ac{4G} \ac{cn}, respectively. %
Regarding the \ac{oai} features, some of them are:

	\begin{itemize}
		\item \textbf{\ac{ue} for \ac{lte} (lte-uesoftmodem):} Support to: \ac{fdd} and \ac{tdd} schemes; bandwidths: 5, 10, and 20~MHz; and \ac{mimo} up to 2x2.\\

		\item  \textbf{\ac{enb} (lte-softmodem):} Support to: \ac{fdd} and \ac{tdd} schemes; bandwidths: 5, 10, and 20~MHz; \ac{mimo} up to 2x2; 256 \ac{qam} modulation in downlink ; and roaming (experimental).

		\item  \textbf{\ac{ue} for \ac{nr} (nr-uesoftmodem):} Support to: \ac{tdd} and \ac{fdd} schemes; subcarrier spacings: 15 and 30~kHz (FR1), 120~kHz (FR2); bandwidths: 10, 20, 40, 80, 100~MHz (273 Physical Resource Blocks); \ac{mimo} up to 4x2; 256 \ac{qam} modulation in downlink.

		\item  \textbf{\ac{gnb} (nr-softmodem):} Support to: \ac{tdd} and \ac{fdd}; subcarrier spacings: 15 and 30~kHz (FR1), 120~kHz (FR2);  bandwidths: 10, 20, 40, 80, 100~MHz (273 Physical Resource Blocks); \ac{mimo} support up to 4x2; 256 \ac{qam} modulation.
	\end{itemize}

In terms of deployment, on the one hand, \ac{srs} offers its documentation with several tutorials, use cases and configuration examples~\cite{ref2}. %
On the other hand, the \ac{oai} has a Gitlab page available at~\cite{ref3} containing some tutorials in the doc folder. %
Concerning this aspect, \ac{srs} is more user friendly. %

Regarding platform compatibility, both run in a linux environment, with \ac{srs} having some precompiled binaries for some distributions such as Ubuntu, openSUSE and Debian. %
It is also possible to compile \ac{srs} in other linux distributions. %
As for \ac{oai}, it has some docker images that can be 	used, but to install it, it is necessary to compile the platform through a script available at \ac{oai}'s Gitlab page. %

Regarding the compatibility of \ac{sdr} devices that can be used together with these platforms, the \ac{oai} is compatible with Eurecom EXMIMO II, \ac{usrp} (B210/X300), BladeRF and LimeSDR for \ac{lte}. %
For \ac{nr} SA, it is compatible with \ac{usrp} (B210/X300). %
Concerning the \ac{srs} in \ac{lte} and \ac{nr} NSA, it is compatible with \ac{usrp} (all models), BladeRF, LimeSDR and also there is compatibility with the \ac{zmq} driver, which simulates a radio virtually inside a machine. %
%All \acp{sdr} compatible with \ac{srs} are capable of running \ac{lte} and \ac{nr}.%

%Regarding the compatibility of \ac{sdr} devices that can be used together with these platforms, for \ac{oai} we have the compatibility of Eurecom EXMIMO II and \ac{usrp} (B210/X300), BladeRF, LimeSDR for \ac{lte} and \ac{usrp} (B210/X300) for \ac{nr}. %
%For \ac{srs}, we have compatibility with \ac{usrp} (all models), BladeRF, LimeSDR and also there is compatibility with the \ac{zmq} driver, which simulates a radio virtually inside a machine. %
%%All \acp{sdr} compatible with \ac{srs} are capable of running \ac{lte} and \ac{nr}.%

	\section{Implementation Methodology} \label{SEC:methodology}

	\begin{figure}[t!]
		\centering
		\includegraphics[width=\linewidth]{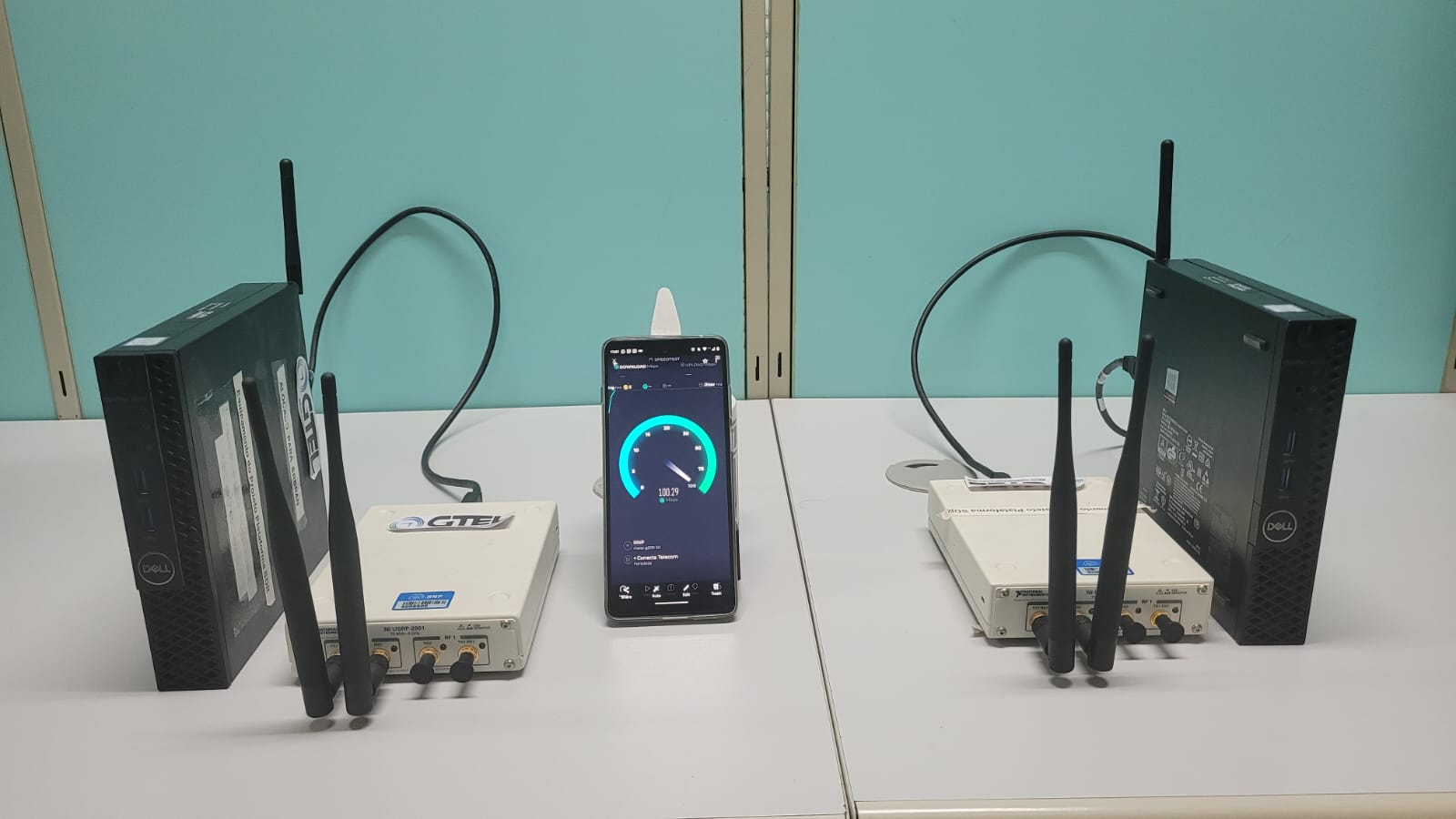}
		\caption{Environment of the deployed SDR network. \ac{ue}'s on the left, \ac{cn} + \ac{gnb} on the right.}
		\label{Fig:testbed}
	\end{figure}

To create a 5G open-source fully-functional network testbed, firstly the environment shown in \FigRef{Fig:testbed} was set up. %
It was composed by 2 mini-PC's Dell OptiPlex 3070 (black boxes in \FigRef{Fig:testbed}), 2 \acp{usrp} B210 (white boxes in \FigRef{Fig:testbed}) and a \textbf{\ac{cots}} [Moto G200 5G] (the smartphone in \FigRef{Fig:testbed}). %

The deployed network testbed had the three main components shown in \FigRef{Fig:netw-arch}. %
One of the mini-PCs acted as \ac{cn} and \ac{gnb}. %
It had access to Internet to allow the \ac{cn} to act as a gateway between the Internet and the private \ac{5G} network created using the platform. %
The other mini-PC was configured as an \ac{ue}. %
Each mini-PC was connected to an \ac{usrp} with two antennas VERT2450. %
Even though not illustrated in \FigRef{Fig:testbed}, a notebook Dell Inspiron 3501 was used to access the mini-PCs through \ac{ssh}. %

Regarding the considered softwares for \ac{cn}, \ac{gnb} and \ac{ue}, some combinations were evaluated. %

First, the \ac{srs} platform was evaluated. %
More precisely, in this implementation the \ac{srs} softwares of \ac{gnb} and \ac{ue}, i.e., \textbf{srsgnb} and \textbf{srsUE}, were considered together with the \textbf{Open5GS \ac{cn}}. %
Besides, in this implementation, the \textbf{\ac{cots}} was also tested with the implemented \ac{5G} private network. %

After that, the \ac{oai} platform was evaluated. %
In this case, the \ac{oai} \ac{5G}/NR softwares, e.i., \textbf{\ac{oai5cn}}, \textbf{nr-softmodem} and \textbf{nr-uesoftmodem}, were used. %
As in the previous case, it was also evaluated the compatibility of a \textbf{\ac{cots}} with the deployed \ac{oai} network. %

Finally, after having completed the platforms individual evaluations, an interoperability test between them was performed. %
First, it was tested the interoperability between the \ac{cn} from Open5GS with the \textbf{\ac{oai} \ac{gnb}}, the \textbf{\ac{oai} \ac{ue}} and the \textbf{\ac{cots}}. %
In the second case, it was tested the interoperability between the  \textbf{\ac{oaicn}} with the \textbf{\ac{srsgnb}}, \textbf{\ac{srsue}} and the \textbf{\ac{cots}}. %

Important to highlight that, while the mini-PC implementing a \ac{ue} used either the \textbf{\ac{srsue}} or the \textbf{\ac{oai} \ac{ue}} softwares, the \textbf{\ac{cots}} [Moto G200 5G] did not used those softwares, since it is already a complete \ac{5G} \ac{ue}. %

%The tests were performed as follows. %
After the connection between \ac{ue} and \ac{gnb} was established, four tests were performed.  %
First, the SpeedTest app was used to evaluate the download rate. % 
After, the iPerf app was also used for similar purposes. % test will also be made after, using termux (a mobile application for emulating terminals). %
Also, a ping was performed from the \ac{ue} to the \ac{cn} to evaluate the latency of the link. %
Finally, a video call was made with Google Meet to qualitatively evaluate the connection. %
%\ac{}: After the connection, a speedtest will be made, 3 times, to ensure the results. %
%An iperf test will also be made after, using termux (a mobile application for
%emulating terminals). %
%In the end, a videocall will be made with another computer, to test the quality of the call. %

For both platforms the following configurations were considered: %
\begin{itemize}

	\item Configuration 1:  40~MHz bandwidth, 46.08~MHz sample rate using 78 band with 3.5~GHz frequency, 30~kHz sub-carrier spacing and \ac{tdd} scheme.

	\item Configuration 2: 20~MHz bandwidth, 23.04~MHz sample rate using 78 band with 3.5~GHz frequency, 30~kHz sub-carrier spacing and \ac{tdd} scheme.
\end{itemize}

We remark that Configuration 1 was not used to evaluate the platforms \acp{ue}. %
It was used to evaluate only the platforms \ac{cn} and \ac{gnb} with the \textbf{\ac{cots}}, since, as already mentioned, the \textbf{\ac{srsue}} only supports bandwidths up to 20~MHz. % 

\section{Results and discussions} \label{SEC:Results}

The installation procedure of the \textbf{srsgnb} was simple. %
An installation guide~\cite{ref4} explaining the installation process was used. %
It describes in a beginner-friendly way how to proceed. %
The \textbf{srsgnb} can be installed on 3 linux system environments, i.e., Ubuntu, Arch Linux and Fedora. %
It is possible to adapt it for other linux environments too. %

As already mentioned, we needed a third-party \ac{cn}, i.e., the \textbf{Open5GS \ac{cn}}. %
However, its installation was harder than the installation of the \textbf{srsgnb}. %
One of the main reasons for this is that the\textbf{ Open5GS \ac{cn}} documentation is harder to follow/understand. %
Furthermore, the documentation focuses on Ubuntu, lacking information regarding the installation in other operational systems. %

The adopted operational system was an Arch Linux, so parts of the installation needed to be changed. %
The most important adjustment made was to create a script file with the processes of the core in Arch Linux. %
Differently from the \textbf{srsgnb}, the \textbf{Open5GS \ac{cn}} installation guide is confusing for beginners. %

After having  installed the \ac{gnb} and the \ac{cn}, the network was ready to be tested by using a \textbf{\ac{cots}} or a third-party \ac{ue}. %
To run the \textbf{\ac{srsgnb}} a configuration file was created, based on the
configuration file already given by the software. %
This file was dependent of the adopted \ac{usrp}. %
For editing this file, a guide for each of the configurations was available at~\cite{ref6}. %
Moreover the lines of the file had comments explaining each parameter, which made it easier to edit it. %

Regarding \ac{oai} platform installation, some difficulties were faced during the installation process. % 
For \textbf{\ac{oai} \ac{gnb}} installation, adjustments were made for the installation of the main dependencies, for the same reason as for the \ac{srs} platform  (the  adaptation for the installation on Arch Linux). %
An example of problem that we faced was during the installation of the libraries \ac{lapack} and \ac{blas}. %
Both of them could not be properly installed. %
To overcome this issue, a solution was to use another library called \ac{openblas}. %
This library is an open-source optimized implementation of \ac{blas} and \ac{lapack}~\cite{ref7}. %
However, the library complete installation (including compilation process) took a long time, i.e., 8~hours. %
For time optimization, the compiled files were recorded in a folder to be reused in possible future re-installations. %

%Also, a ``build$\_$helper'' file was modified to create an installation routine for the Arch Linux environment. %
%For \ac{oai5cn}, another fix was, while building the \ac{gnb}, to track the output text file, which enabled installation with administrator privileges. %

In \ac{oai}, installing the \ac{cn} was easier than installing the \ac{gnb}. %
No documentation was found regarding the \ac{gnb} installation procedure. %
We faced similar issues during the \textbf{\ac{oai} \ac{ue}} setup. %
The documentation is a little confusing sometimes, since they are all placed in a repository~\cite{ref3}, and they require more time to assemble and understand. %
Also, some configuration files are coded in C, which means that, to change them, you have to recompile the platform, which takes time. %

Regarding the tests with the platforms \acp{ue} and the \textbf{\ac{cots}}, both platforms performed worse with their own \acp{ue} compared to using \textbf{\ac{cots}}. %
More specifically, configuring the \textbf{\ac{cots}} was easier since we only needed to configure a \ac{sim} card and add its \ac{imsi} number to the \ac{cn}. %
Moreover, the connection between the \textbf{\ac{cots}} and the \ac{gnb} was also more stable than the connection between an \ac{ue} of a given platform and its own \ac{gnb}. %
This can be partially explained by the fact that the \textbf{\ac{cots}} hardware was superior to the one used with the \textbf{\ac{srsue}} and \textbf{\ac{oai} \ac{ue}} softwares. %
%For both platforms, using their \ac{ue} softwares was harder than directly using the \textbf{\ac{cots}}. %
%More bugs, worst performance, and even connection problems between \ac{ue} and \ac{gnb} from the same platform were detected. %
Thus, even though we tested the platforms \acp{ue}, the \textbf{\ac{cots}} was used to obtain the network \acp{kpi} presented in the following. %

\begin{figure}[!t]
	\centering
	\subfloat[VideoCall with Google Meet.]{%
		\includegraphics[scale = 0.145]{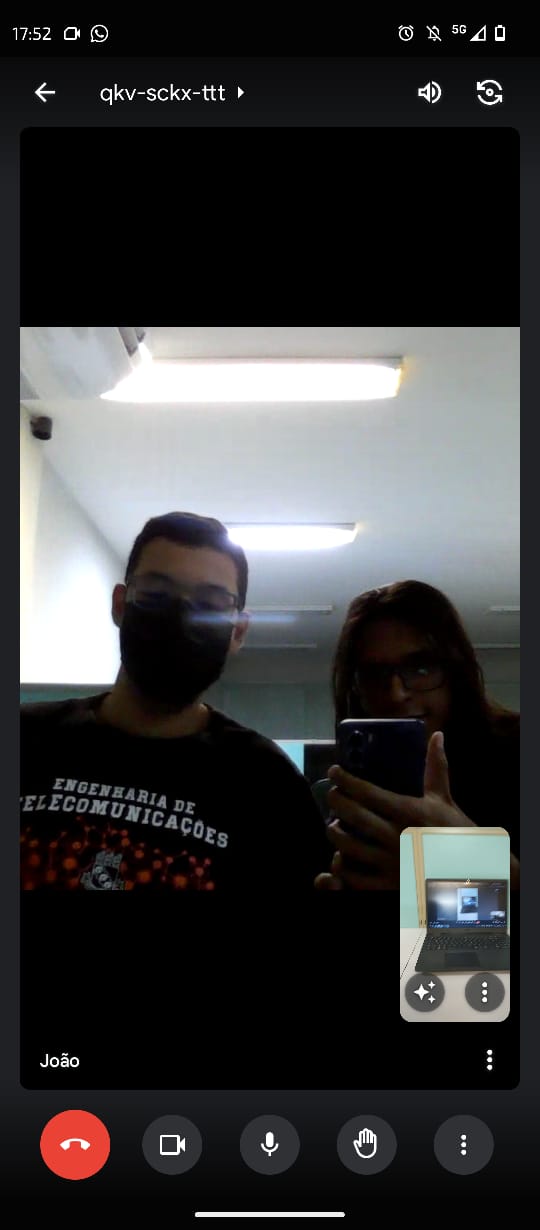}
		\label{FIG:down-config-two-srsproj}
	}
	\subfloat[SpeedTest.]{ % - Config.01 for \ac{oai}
		\includegraphics[scale = 0.145]{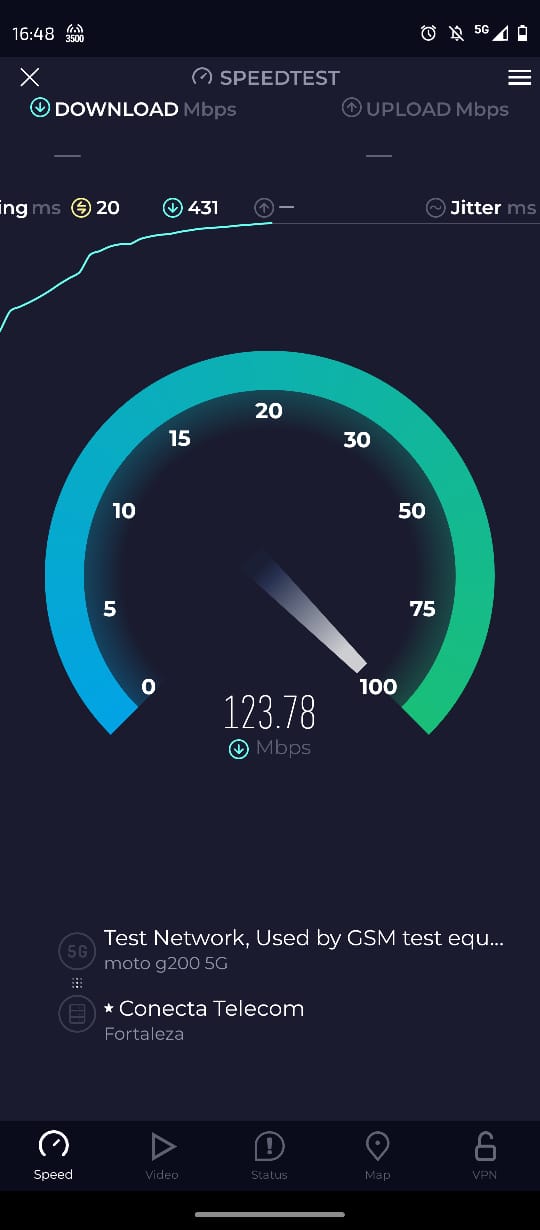}
		\label{FIG:down-config-one-srsproj}
	}
	\caption{Screenshots of the tests with Google Meet and SpeedTest, both conducted with \ac{oai} configuration~1.}\label{FIG:Architecture}
	\vspace*{-0.3cm}
\end{figure}

\FigRef{FIG:Architecture} presents screenshots of Google Meet and SpeedTest during the tests. %
Regarding the video call quality, illustrated in \FigRef{FIG:down-config-two-srsproj}, on the one hand, with \ac{oai}, for a distance between \ac{ue} and \ac{gnb} of up to 10~meters, the video call had a good quality with a sharp image and without lags or freezing. %
%This good quality  
%  The coverage area is about 10 meters, approximately. In the environment
%  described in \cite{ref1}, it works successfully, as can be seen in Fig.11.
On the other hand, with \ac{srs}, the video call did not have a good quality. %
The connection was unstable with the image frequently freezing. %

Concerning the tests with SpeedTest, \TabRef{TAB:Download} presents the average download rate achieved by the \textbf{\ac{cots}} in each evaluated setup presented in \SecRef{SEC:methodology}. %
These download rates are average of the download rates measured by the SpeedTest. %

Analyzing the results presented in~\TabRef{TAB:Download}, we can see that for a given configuration and \ac{gnb}, the download rate is similar for both \acp{cn}. %
The main difference appears when comparing \acp{gnb} of different platforms. %
This can be explained by the fact that \textbf{\ac{oai} \ac{gnb}} supports 256~\ac{qam}, while the \textbf{\ac{srsgnb}} only supports up to 64~\ac{qam}. %

%\begin{center} 
  \begin{table}[tbp]
  	\center
    \caption{Average download rate achieved by the \textbf{\ac{cots}} in each evaluated setup, measured by SpeedTest.} \label{TAB:Download}
  \begin{tabular}{| c | c | c | c |}
    \hline
    Configuration & \ac{cn} & \thead{\ac{oai} \ac{gnb} } & \thead{ \ac{srs} \ac{gnb}} \\
    \hline
    Configuration 1 & \ac{oaicn} & 123.78 Mbps & 59.61 Mbps \\
    \hline
    Configuration 1 & Open5GS & 116.19 Mbps & 62.69 Mbps \\
    \hline
    Configuration 2 & \ac{oaicn} & 52.04 Mbps & 17.70 Mbps\\
    \hline
    Configuration 2 & Open5GS & 52.63 Mbps & 17.24 Mbps \\
    \hline
  \end{tabular}
  \end{table}
%  \end{center}

\begin{figure}[t!]
	\centering
	\includegraphics[width=0.9\linewidth]{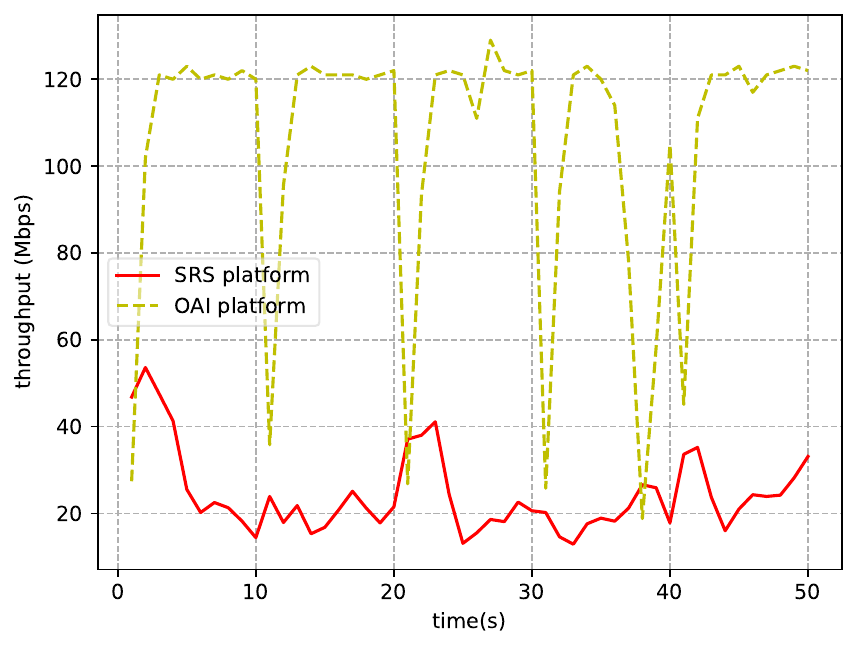}
	\caption{Example of throughput temporal evolution measured by iPerf.}
	\label{FIG:throughput}
\end{figure}

Regarding the tests with iPerf, \FigRef{FIG:throughput} presents an example of the temporal evolution of the throughput for both \ac{sdr} platforms with configuration 1, measured by iPerf. %
The throughput was measured at each second. %
The difference between the throughput achieved with each one of the platforms is highlighted. %
As already mentioned, it was mainly due to the difference of the
\ac{qam} order used.  The downward cycle of the throughput of the OAI platform, shown in \FigRef{FIG:throughput}, can be explained by the difference in the data capture threshold and the GNB output threshold, as the GNB output being lower than the data capture.  %

%Similarly, \FigRef{FIG:throughput} presents the throughtput rate of both networks at configuration 1 in y axis, \ac{srs} Project with
%Open5GS and \ac{oai} with \ac{oai5cn}, for 50 samples in x axis. %
%The throughput was measured within 1 second per sample. % 

%In \FigRef{FIG:throughput}, the difference between the througput achieved with each one of the platforms is highlighted. %
%It is also possible to see the difference caused by
%\ac{qam} order used.%

	\begin{figure}[t!]
	\centering		
	\subfloat[Complete Y-axis.]{%
		\includegraphics[width=0.9\linewidth]{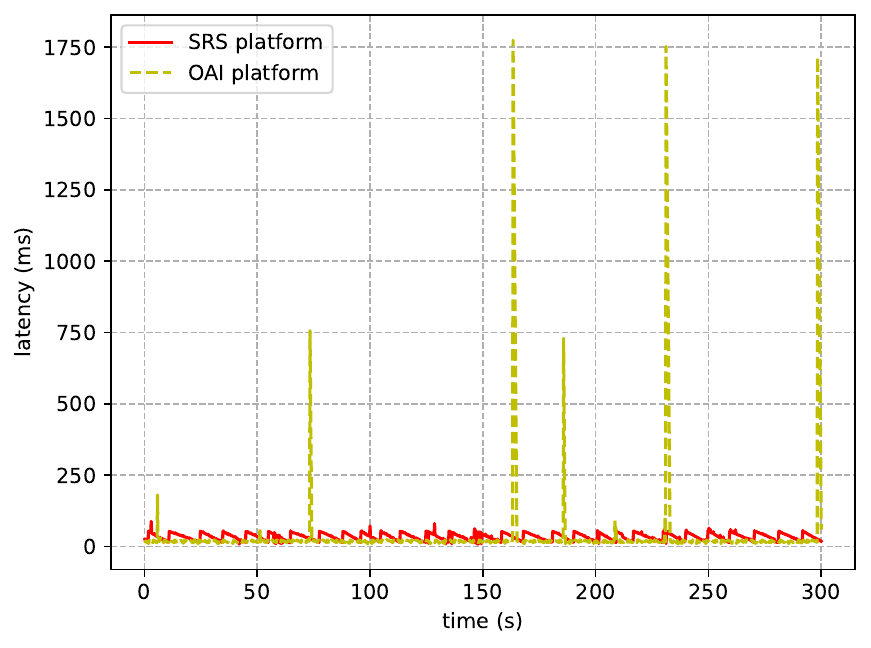}
		\label{FIG:latency_complete}
	}
	
	\subfloat[Zoom in Y-axis.]{%
		\includegraphics[width=0.9\linewidth]{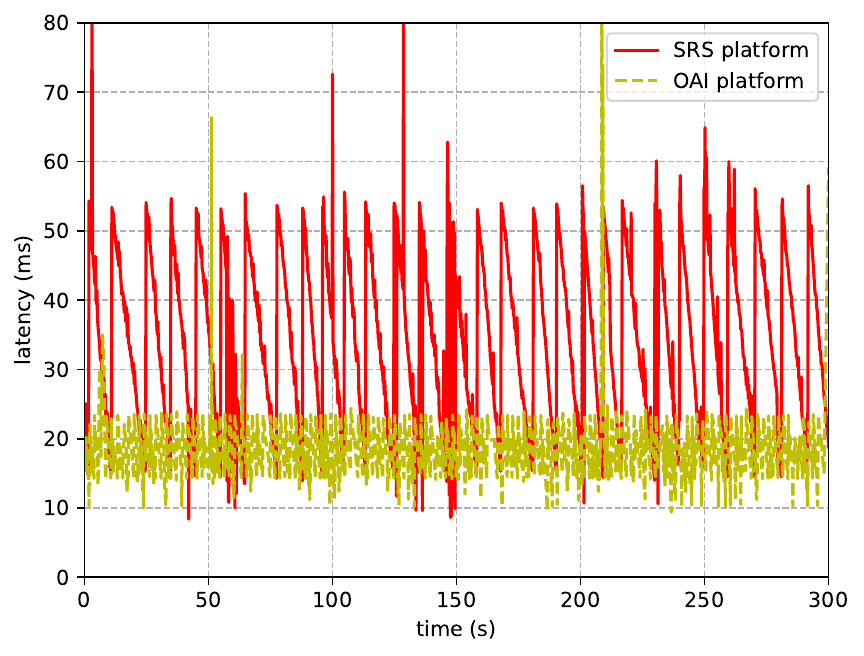}
		\label{FIG:latency_zoom}
	}
	\caption{Example of latency temporal evolution for both platforms with configuration~1.}
	\label{FIG:latency}
\end{figure}

Similarly, \FigRef{FIG:latency} presents the temporal evolution of the latency for both \ac{sdr} platforms with configuration 1. %
Figs.~\ref{FIG:latency_complete} and~\ref{FIG:latency_zoom} refer to the same case, the difference is that \FigRef{FIG:latency_zoom} presents a zoom in the Y-axis. %
%,\ac{srs} Project with
%Open5GS and \ac{oai} with \ac{oai5cn}, for 300 seconds in x axis. %
The measurements were performed at each 0.2~seconds. %
Comparing \ac{srs} and\ac{oai} in \FigRef{FIG:latency_zoom}, we can see the the \ac{oai} average value is lower than the \ac{srs} average value. %
However, the \ac{oai} presented spikes in \FigRef{FIG:latency_complete}. %
The spikes were due to a refresh of the network, instructed by the \ac{gnb} to occur approximately at each 75~seconds. These spikes can be explained by a configuration made by the developers of the OAI platform, so that the GNB can have its memory cache cleared and store new network information and perform operations, since the platform is coded in order to perform recompilations while running. %

Also, notice in \FigRef{FIG:latency}, the sawtooth pattern of the \ac{srs} curve. %
This can be explained by the fact that, in \ac{srs} Project v.23.3, some \ac{rrc}
procedures are not implemented, such as connection reestablishment \cite{ref9}. %
The lack of these  procedures have as consequence the sawtooth pattern observed in the latency curve. %
More precisely, after a certain period without activity, the \textbf{\ac{cots}} entered in idle mode. %
When traffic arrived to it, it needed to initiate a new connection, which takes more time than just performing a connection reestablishment procedure, which was not available. %

	\section{Conclusions and future perspectives} \label{SEC:Conclusion}

The present paper presented our main findings related to our experience in deploying and using an experimental \ac{5G} testbed. %
Two different \ac{sdr} platforms were evaluated, i.e., the \ac{srs} and the \ac{oai}. %

Overall, we were able to use both of them and they worked with a \textbf{\ac{cots}}, which means that researchers and developers can use both of them to deploy their own \ac{5G} testbeds that can be used for different purposes. %

More specifically, based on our evaluations, the \ac{srs} fits better for beginners. % will probably fulfill all the expectations. %
It is easier to use, more intuitive and have a better documentation. %
%, and if a user don't have the hardware needed, he can use \ac{zmq} to simulate the connection and receive the metrics.
However, for researchers and developers seeking for a more complete platform, \ac{oai} is probably the best choice. %
Its learning curve takes more time, but rewards the developer seeking for a more flexible network. %

As a future perspective of this work, one can use the \ac{5G} testbed that was deployed to implement applications that can explore the advantages of a private \ac{5G} network. %
For example, connecting \ac{5G} modules to security cameras in order to transmit in real time to a central unit the images that will be used as input of a \ac{ml} solution to detect undesired events. %
Another possibility is using the \ac{5G} testbed to evaluate solutions that improves the network itself, for example to test radio resource management algorithms. %

    \begin{acronym}[PCNA]
		\acro{4G} [4G] {Fourth Generation}
		\acro{5G} [5G] {Fifth Generation}
		\acro{6G} [6G] {Sixth Generation}
		\acro{sim} [SIM] {Subscriber Identity Module}
		\acro{srs} [srsRAN] {Software Radio Systems Radio Access Networks}
		\acro{srsp} [srsRAN Project] {Software Radio Systems Radio Access Networks Project}
		\acro{ue} [UE] {User Equipment}
		\acro{srsue} [srsUE]  {Software Radio Systems User Equipment}
		\acro{srsenb} [srsENB] {Software Radio Systems eNodeB}
		\acro{srsepc} [srsEPC] {Software Radio Systems Evolved Packet Core}
		\acro{srsgnb} [srsENB] {Software Radio Systems gNodeB}
		\acro{epc} [EPC] {Evolved Packet Core}
		\acro{ran} [RAN] {Radio Access Network}
		\acro{cn} [CN] {Core Network}
		\acro{enb} [eNB] {eNodeB}
		\acro{gnb} [gNB] {gNodeB}
		\acro{lte} [LTE] {Long Term Evolution}
		\acro{nr} [NR] {New Radio}
		\acro{nsa} [NSA] {Non-Standalone}
		\acro{sa} [SA] {Standalone}
		\acro{fdd} [FDD] {Frequency Division Duplex}
		\acro{tdd} [TDD] {Time Division Duplex}
		\acro{mimo} [MIMO] {Multiple Input Multiple Output}
		\acro{oaicn} [OAI-CN] {OpenAirInterface-Core Network}
		\acro{oai5cn} [OAI-5GCN] {OpenAirInterface 5G Core Network}
		\acro{sdr} [SDR] {Software-Defined Radio}
		\acro{ssh} [SSH] {Secure Shell Protocol}
		\acro{oai} [OAI] {OpenAirInterface}
		\acro{oaignb} [OAI gNB] {OpenAirInterface gNodeB}
		\acro{oaiue} [OAI nrUE] {OpenAirInterface New Radio User Equipment}
		\acro{zmq} [ZMQ] {ZeroMQ}
		\acro{usrp} [USRP] {Universal Software Radio Peripheral}
		\acro{bs} [BS] {Base Station}
		\acro{cots} [COTS UE] {Commercial Off-The-Shelf User Equipment}
		\acro{imsi} [IMSI] {International Mobile Subscriber Identity}
		\acro{open} [O5GS] {Open5GS}
		\acro{cblol} [CBLOL] {Brazilian League of Legends Championship}
		\acro{blas} [BLAS] {Basic Linear Algebra Subprograms}
		\acro{lapack} [LAPACK] {Linear Algebra Package}
		\acro{openblas} [OpenBLAS] {Open Basic Algebra Subprograms}
		\acro{kpi} [KPI] {Key Performance Indicator}
		\acro{sim} [SIM] {Subscriber Identity Module}
		\acro{ml} [ML] {Machine Learning}
    	\acro{qam} [QAM] {Quadrature Amplitude Modulation}
    	\acro{rrc} [RRC] {Radio Resource Control}
	\end{acronym}


\begin{thebibliography}{99}
	\bibitem{ref0} K. Kaur, V. Mangat and K. Kumar, ``Architectural Framework, Research Issues and Challenges of Network Function Virtualization'', \textit{2020 8th International Conference on Reliability, Infocom Technologies and Optimization (Trends and Future Directions) (ICRITO)}, Noida, India, 2020, pp. 474-478, doi: 10.1109/ICRITO48877.2020.9197802.
	\bibitem{refb} S. Katz and J. Flynn, ``Using software defined radio (SDR) to demonstrate concepts in communications and signal processing courses,'' \textit{2009 39th IEEE Frontiers in Education Conference}, San Antonio, TX, USA, 2009, pp. 1-6, doi: 10.1109/FIE.2009.5350716.
	\bibitem{refc} Wilker de O. Feitosa, Ruan A. da Silva, Victor F. Monteiro and Francisco R. P. Cavalcanti, ``RSRP Prediction on LTE Network Testbed Using a Software Defined Radio (SDR) Platform'', \textit{XL Simpósio Brasileiro de Telecomunicações e Processamento de Sinais (SBrT2022)}, Santa Rita do Sapucai, Brazil, 2022, doi: 10.14209/sbrt.2022.1570814230. 
	\bibitem{ref5} SRSRAN PROJECT, ``Running srsRAN Project'', Software Defined
	Radio Radio Acess Network Project(srsRAN Project). Endereço:
	\url{https://docs.srsran.com/projects/project/en/latest/user_manuals/source/running.html#manual-running}
	(Acesso em: 15/05/2023).
	\bibitem{ref2} SRS, ``Feature List'', Software Radio System(SRS). Endereço: \url{https://docs.srsran.com/projects/project/en/latest/general/source/2\_feature\_list.html} (Acesso em 05/06/2023).
	\bibitem{ref3} OAI, ``OpenAirInterface documentation overview'', Open Air Interface(OAI). Endereço: \url{https://gitlab.eurecom.fr/oai/openairinterface5g/-/tree/develop/doc} (Acesso em 05/06/2023).
%	\bibitem{ref1} 3GPP, ``Study on Channel Model for Frequencies from 0.5 to
%		100GHz.'', 3rd Generation Partinership Project (3GPP), TR 38.901, set. de
%		2017, v14.2.0. Endereço: \url{http://www.3gpp.org/DynaReport/38901.htm} (Acesso em 05/06/2023).
%		1986.
    \bibitem{ref4} SRSRAN PROJECT, ``Installation Guide'', Software Defined
      Radio Radio Acess Network Project(srsRAN Project). Endereço:
      \url{https://docs.srsran.com/projects/project/en/latest/user_manuals/source/installation.html#manual-installation}
      (Acesso em: 15/05/2023).
    \bibitem{ref6} SRSRAN PROJECT, ``Configuration Reference'', Software Defined
      Radio Radio Acess Network Project(srsRAN Project). Endereço:
      \url{https://docs.srsran.com/projects/project/en/latest/user_manuals/source/config_ref.html}
      (Acesso em 15/05/2023).
    \bibitem{ref7} OPENBLAS, ``OpenBLAS Wiki'', Open Basic Linear Algebra
      Subprograms(OpenBLAS). Endereço:
      \url{
      https://github.com/xianyi/OpenBLAS/wiki}
      (Acesso em 15/05/23).
%    \bibitem{ref8} SRSRAN 4G, ``Configuration Reference'', Software Defined
%      Radio Radio Acess Network 4G(srsRAN 4G). Endereço:
%      \url{      https://docs.srsran.com/projects/4g/en/latest/feature_list.html}
%      (Acesso em 15/05/2023).
    \bibitem{ref9} SRSRAN PROJECT, ``Release Notes'', Software Defined
      Radio Radio Acess Network Project(srsRAN Project). Endereço:
      \url{https://docs.srsran.com/projects/project/en/latest/general/source/5_release_notes.html}
      (Acesso em 15/05/2023).
    \end{thebibliography}
\end{document}